\input phyzzx
\sequentialequations
\overfullrule=0pt
\tolerance=5000
\nopubblock
\twelvepoint

\line{\hfill Caltech 68-2032, IASSNS 96/04}
\line{\hfill hep-ph/9601279 }
\line{\hfill January 1996}

\REF\ew{See F. Merritt, H. Montgomery,
A. Sirlin, and M. Swartz, {\it Precision Tests of Electroweak Physics:
Current Status and Prospects for the Next Two Decades} in
{\it Particle
Physics: Perspectives and Opportunities}, ed. Peccei {\it et al}., 
(World Scientific, 1995) and references therein.}

\REF\susyun{See S. Dimopoulos, S. Raby, and F. Wilczek, 
{\it The Unification of Couplings}, Physics 
Today {\bf 44}, October 1992, p. 25, and references therein.}

\REF\cdf{A. Bhatti, {\it Inclusive Jet Production at Tevatron}, 
Rockefeller-CDF preprint /CDF/JET/PUBLIC/3229, June 1995.} 

\REF\bbp{V. Barger, M. Berger, and R. Phillips, {\it Thresholds in
$\alpha_s$ Evolution and the $p_T$ Dependence of Jets},
hep-ph/9512325,
December 1995.}

\REF\wein{See especially S. Weinberg, {\it New Approach to the
Renormalization Group}, Phys. Rev. {\bf D8}, 3497 (1973).} 

\FIG\diag{One loop diagrams contributing to four quark processes in the
QCD sector of the MSSM.  Lines corresponding to squarks and gluinos are
labelled, as are the Lorentz and color indices of the external quarks.  
The remaining diagrams may be obtained from the above by 
permuting indices.}

\FIG\diag{One loop diagrams contributing to two quark - two gluon processes.
The independent external momenta are labelled, where we use
 the convention that positive momentum for the quarks goes in the same 
direction as the charge flow, and for gluons it goes into the vertex.  }
 
\titlepage
\title{QCD Interference Effects of Heavy Particles Below Threshold}

\author{Per Kraus\foot{Research supported in part by a DuBridge Fellowship
and by DOE grant DE-FG03-92-ER40701}}
\centerline{{\it Department of Physics }}
\centerline{{\it Lauritsen Laboratory}}
\centerline{{\it California Institute of Technology}}
\centerline{{\it Pasadena, Cal. 91125 }}
\vskip .2cm
\author{Frank Wilczek\foot{Research supported in part by DOE grant
DE-FG02-90ER40542.~~~wilczek@sns.ias.edu}}
\vskip.2cm
\centerline{{\it School of Natural Sciences}}

\centerline{{\it Institute for Advanced Study}}
\centerline{{\it Olden Lane}}
\centerline{{\it Princeton, N.J. 08540}}
 
\endpage
 
\abstract{We consider how two classes of heavy particles: extra
vector-like families, and strongly interacting superpartners, manifest
themselves below threshold, by interference of virtual loops with
normal
QCD processes.  Quantitative estimates are presented.}
 
\endpage
 
\chapter{Introduction}

%detecting heavies below threshold: historical

%`constraint' from unification of couplings

Although the Standard Model has proven to be remarkably successful at
presently accessible energies, there is universal agreement that it is
incomplete -- that new structure will emerge at higher energies.  
The existence of new heavy particles can be 
inferred even at 
energies far below threshold
if they mediate processes which violate symmetries of the low
energy theory.  A classic example is the weak interactions, beginning
with Fermi's theory of $\beta$ decay described by a Lagrangian with symmetry 
breaking four fermion operators, and culminating in the discovery of W and
Z.  Even if it does not have the qualitative effect of
violating a symmetry, exchange of   
virtual heavy particles can in principle generate observable
quantitative
consequences.  This has been
much discussed in the context of precision electroweak measurements
[\ew ].  
Here we will briefly address another possibility, which potentially yields
much larger effects, though unfortunately with more poorly controlled
backgrounds.  That is, we will  consider the effects of  
interference between 
conventional QCD amplitudes and amplitudes involving virtual exchange
of various possible heavy particles.

Given the apparent unification of couplings, which is at least
in good semi-quantitative in agreement with the minimal supersymmetric
Standard Model (MSSM) [\susyun ], 
most interest attaches to a rather short list
of particles -- those whose existence would not render this success
coincidental.  Besides
the minimal superpartners, 
the catalog of additional particles which can appear at
the TeV scale is quite short; essentially, one must consider
complete
SU(5) multiplets.  
We shall mainly focus on the superpartners in
the MSSM, but we shall compare the (more easily calculated) effect of 
additional vectorlike families and briefly mention new
Z$^\prime$ 
gauge bosons.  For simplicity, we will restrict our attention to processes
involving four external quarks.   

\chapter{Vacuum Polarization; Vector Families}

%effects of new families

The simplest QCD interference effects arise from vacuum polarization
due to heavy particles.  We write the full gluon propagator, 
$D'^{ab}_{\mu \nu}$, as
$$
iD'^{ab}_{\mu \nu}~=~iD^{ab}_{\mu \nu} + iD^{am}_{\mu \alpha} 
(i\Pi^{\alpha \beta}_{mn}) iD^{nb}_{\beta \nu} + \dots
$$
and
$$
i\Pi^{\alpha \beta}_{mn}\!(q)~=~i\delta_{mn}(q^{\alpha}q^{\beta}-g^{\alpha
\beta}q^2)i\Pi\!(q^2)~.
$$
We renormalize by zero momentum subtraction; that is, by computing
$\Pi\!(q^2)-\Pi\!(0)$.  Well below threshold it is appropriate to expand the 
result in powers of $q^2/m^2$, and to keep only the lowest term.  
For scalars of mass $m_s$ in representation R
of color SU(3) we then have:
$$
\Pi\!(q^2)-\Pi\!(0)~=~{\alpha_s \over 120\pi}{q^2 \over m_s^2}{\rm T(R)}
~+~{\rm O}\!\left({q^4\over m^4_s}\right) ~,
$$
where T(R) is defined by Tr$(T_R^a T_R^b)~=~$T(R)$\delta^{ab}$.
For a Dirac fermion:
$$
\Pi\!(q^2)-\Pi\!(0)={\alpha_s \over 15\pi}{q^2 \over m_f^2}{\rm T(R)}
~+~{\rm O}\!\left({q^4 \over m^4_f}\right)~.
$$
A Majorana or Weyl fermion contributes one half of this amount.

Interference of vacuum polarization diagrams with one gluon exchange
diagrams gives the change in cross section:
${\Delta \sigma \over \sigma}~=~-2[\Pi\!(q^2)-\Pi\!(0)].$
Therefore, $n_f$ families of Dirac fermions contribute
$$
{\Delta \sigma \over \sigma}~=~-{2\alpha_s \over 15\pi}{q^2 \over m_f^2}
n_f {\rm T(R)}~.
$$
This formula is roughly valid until $|q^2/4m_f^2|=1$, at which
point the cross section for the dominant
t and u channel processes has increased by
$\Delta \sigma/\sigma \approx 2n_f$ T(R) \%, taking $\alpha_s 
\approx 1/8$.  Strangely enough, the s channel process is actually
suppressed -- a warning that one must not think too naively about
timelike
{\it versus\/} spacelike effective couplings.  
For $n_f$ vectorlike supersymmetric families 
(SU(5) $\bar {10} + 5 + 10 + \bar 5$), assumed mass-degenerate, 
one has in the same approximations
$$
{\Delta \sigma \over \sigma}~=~ -{\alpha_s \over 6\pi}{q^2 \over m_f^2}
n_f {\rm T(R)}~ \sim 5 n_f\%
\eqn\vectorfamily
$$
(${\rm T(R)} ~=~ 4~{\rm triplets} \times {1\over 2} = 2$).  
There is of course no difficulty in using the accurate vacuum
polarization formulas.  This gives a slightly smaller value for the
correction at $|q^2/4m| = 1$, and turns over at large $q^2$ into
the familiar logarithmic running of the coupling.

\chapter{Supersymmetry}

%effects of supersymmetric particles

In the MSSM, the particles contributing to the vacuum polarization are
Majorana gluinos in the adjoint representation, and 6 flavors $\times$ 2
chiralities $~=~$ 12 squarks in the fundamental representation.  
So, setting
all masses to to be equal,
$$
\Pi\!(q^2)-\Pi\!(0)={3 \alpha_s \over 20 \pi}{q^2 \over m^2}~.
$$
However in this case
vacuum polarization  is not the whole story, since
the presence of gluino-squark
Yukawa couplings leads to additional contributions from
vertex corrections and box diagrams.  Supersymmetry dictates that
the strength of these couplings
have fixed numerical ratios to the ordinary strong coupling, and 
these diagrams are in no sense negligible.  We have evaluated the relevant  
diagrams, which are displayed in Fig. 1, in Feynman gauge.  
In addition we  have evaluated some of
the diagrams corresponding to two quark - two gluon processes, which are
displayed in Fig. 2.  More nonvanishing diagrams can be obtained from
those shown by permuting Lorentz and color indices.  
Although there would seem to
be  diagrams containing two gluon - two squark vertices which
should be included with the diagrams in Fig. 2,  these turn out to vanish
for on shell, massless quarks and so can be omitted.  
In each of the diagrams we
can choose the squarks to be of either L or R type; in our results below
we have added the corresponding amplitudes together.   
 Similarly, for the Fig. 2 diagrams we have added the amplitude
for the displayed diagram plus the diagram with gluon labels exchanged.
For simplicity, we have given all the squarks and gluinos a common mass, $m$.
The results for on shell, massless quarks and gluons are summarized below. 

\settabs\+&1a\qquad&\cr
\+\cr

\+&1a: &$-{i\alpha_{s}^{2}\over 72 m^2}~(T_a)_{ij}~(T_a)_{mn}~
    (\gamma^\mu)_{\alpha\beta}~(\gamma_\mu)_{\gamma\delta}$\cr
\+\cr

\+&1b: &$ {3i\alpha_{s}^{2}\over 8 m^2}~(T_a)_{ij}~(T_a)_{mn}~
    (\gamma^\mu)_{\alpha\beta}~(\gamma_\mu)_{\gamma\delta}$\cr

\+\cr

\+&1c: &$ -{i\alpha_{s}^{2}\over 6 m^2}~(T_a T_b)_{in}~(T_b T_a)_{mj}
    \left[ (\gamma^{\mu})_{\alpha\beta}~(\gamma_\mu)_{\gamma\delta}~
    +~ (\gamma^{\mu}\gamma_5)_{\alpha\beta}~(\gamma_{\mu}\gamma_5)_{\gamma
     \delta}~\right.
$\cr\smallskip
\+&&$\left.\qquad\qquad\qquad\qquad\qquad\qquad
    -~2~(1)_{\alpha\beta}~(1)_{\gamma\delta}~
     -~2~(\gamma_5)_{\alpha\beta}~(\gamma_5)_{\gamma\delta}\right]$\cr
\+\cr

\+&1d: &${i\alpha_{s}^{2}\over 6 m^2}~(T_a T_b)_{ij}~(T_a T_b)_{mn}
    \left[ (\gamma^{\mu}C)_{\gamma\alpha}~(C^{-1}\gamma_\mu)_{\beta\delta}~
    +~ (\gamma^{\mu}\gamma_5C)_{\gamma\alpha}~
    (C^{-1}\gamma_{\mu}\gamma_5)_{\beta\delta}~\right.
$\cr\smallskip
\+&&$\left.\qquad\qquad\qquad\qquad\qquad\qquad
    -~2~(C)_{\gamma\alpha}~(C^{-1})_{\beta\delta}~
     -~2~(C\gamma_5)_{\gamma\alpha}~(C^{-1}\gamma_5)_{\beta\delta}\right]$\cr

\+\cr

\settabs\+&2a\qquad&\cr

\+&2a: &${\alpha_{s}^{2}\over 6m^2}~f_{eac}~f_{ebd}~(T_{c}T_d)_{ij}
\left[\gamma^{\mu}(3p_{1}^{\nu}-2p_{2}^{\nu})~+~(-p_{1}^{\mu}+2p_{2}^{\mu})
\gamma^{\nu}\right.$\cr
\smallskip
\+&&$\left.\qquad\qquad\qquad\qquad\qquad\qquad
+~\not\!{p_3}\gamma^{\mu}\gamma^{\nu}~-~1/2(\gamma^{\mu}
\!\!\!\not\!{p_3}
\gamma^{\nu}-\gamma^{\nu}\!\!\!\not\!{p_3}\gamma^{\mu})\right]$\cr
\smallskip
\+&&$+~~ \left\{~ \mu \leftrightarrow \nu \quad ;\quad a\leftrightarrow b \quad ; \quad
       p_{3}\rightarrow p_{1}-p_{2}-p_{3}~\right\}$\cr

\+\cr

\+&2b:&$-{\alpha_{s}^{2}\over 12m^2}~f_{bcd}~(T_{c}T_{a}T_{d})_{ij}
  \left[2\gamma^{\mu}(p_{1}^{\nu}+p_{2}^{\nu})~-~6(p_{1}^{\mu}+p_{2}^{\mu})
   \gamma^{\nu}~+~4\!\!\!\not\!{p_3}\gamma^{\nu}\gamma^{\mu}\right]$\cr

\smallskip
\+&&$+~~ \left\{~ \mu \leftrightarrow \nu \quad ;\quad a\leftrightarrow b \quad ; \quad
       p_{3}\rightarrow p_{1}-p_{2}-p_{3}~\right\}$\cr

\+\cr

\+&2c:&${i\alpha_{s}^{2}\over 6m^2}~(T_{c}T_{a}T_{b}T_{c})_{ij}
  \left[\gamma^{\mu}(2p_{1}^{\nu}-3p_{2}^{\nu}-p_{3}^{\nu})~+~
(-2p_{1}^{\mu}+p_{2}^{\mu})\gamma^{\nu}~-~\!\!\!\not\!{p_3}g^{\mu\nu}\right]
$\cr
\smallskip
\+&&$+~~ \left\{~ \mu \leftrightarrow \nu \quad ;\quad a\leftrightarrow b \quad ; \quad
       p_{3}\rightarrow p_{1}-p_{2}-p_{3}~\right\}$\cr

\+\cr

\+&2d:&${\alpha_{s}^{2}\over 30 m^2}~f_{abc}~(T_c)_{ij}
\left[-2\gamma^{\mu}p_{3}^{\nu}~+~2(p_{1}^{\mu}-p_{2}^{\mu})\gamma^{\nu}
~+~{p_{1}^{\mu}p_{3}^{\nu}-p_{2}^{\mu}p_{3}^{\nu}\over 
p_{3}^{\sigma}(p_{1\sigma}-p_{2\sigma})} \!\!\not\!{p_3}
~+~g^{\mu\nu}\!\!\not\!{p_3}
\right]$\cr

\+\cr

\+&2e:&${\alpha_{s}^{2}\over10 m^2}~f_{abc}(T_c)_{ij}
  \left[-8\gamma^{\mu}p_{3}^{\nu}+8(p_{1}^{\mu}-p_{2}^{\mu})\gamma^{\nu}
~-~{p_{1}^{\mu}p_{3}^{\nu}-p_{2}^{\mu}p_{3}^{\nu}\over 
p_{3}^{\sigma}(p_{1\sigma}-p_{2\sigma})} \!\!\not\!{p_3}~ +~
9g^{\mu\nu}\!\!\not\!{p_3}
\right]$\cr
\+\cr

$f_{abc}$ are SU(3) structure constants, and $T_a$ are SU(3)
matrices in the fundamental representation.
$C$ is the charge conjugation operator, which arises due to the Majorana
fermion propagators.  When contracting with external spinors the following
formulas are useful: $C\bar{u}^{T}(p,\pm)=v(p,\pm)\quad;\quad
u^{T}(p,\pm)C^{-1}=-\bar{v}(p,\pm)$.  In the results for the diagrams of Fig.
2, we have suppressed the spinor indices; all gamma matrices have the spinor
indices $(\gamma^{\mu})_{\alpha\beta}$.

  In Feynman gauge 
the dominant contribution to four quark processes comes from the vertex
correction involving a triangle with two gluinos and one squark.  This
diagram contributes
$$
{\Delta \sigma \over \sigma}~=~{3 \alpha_s \over 8\pi}{q^2 \over m^2}~.
$$
Note that this comes with the opposite sign as compared to the vacuum
polarization piece. Combining these, we find a small {\it decrease\/} 
in the 
cross section for t and u channel processes, of order 
$\Delta \sigma/\sigma \approx -1  $ \% at $q^2/4m^2=-1$.  Insofar as
this is representative of the overall magnitude of the effect, it is
disappointingly small.  

Let us note
that 
although the box diagrams are numerically smaller they do lead to a 
different angular distribution.  

\chapter{Comments}

%new vector mesons in crossed channels

%overall magnitude

1. Recent CDF data on inclusive jet production in the transverse momentum
region $200 <p_T<420 $ GeV indicates a cross section apparently exceeding
QCD predictions by several tens of percent [\cdf ].  
The results of the present paper indicate
that the data cannot readily be explained by a mass threshold in the
minimal
supersymmetric standard model,
as the main calculated effect is too small and of the wrong sign.
Extra vector-like sumpersymmetric families will give a positive
effect, 
but still rather small unless there are many such families.  
After our calculations were completed, but before this
note was prepared a paper by Barger, Berger, and Phillips [\bbp ] appeared, in
which they
claim that a mass threshold at 200 GeV, with
sufficient particle content to  turn off the running of
$\alpha_s$ asymptotically, 
can lead to a substantial (20 \%) increase in jet production
at 500 GeV.  They do not consider the effect of
Yukawa couplings, and so their estimates might apply for generic extra 
strongly-interacting matter,
but
not for the
superpartners of ordinary matter.  

2.  Above we have essentially 
considered corrections to quark-quark scattering
due to virtual 
heavy particles; one should also consider whether exchange of such
particles
affects the amplitudes for finding quarks inside the initial
projectiles, {\it i. e}. the structure functions.  Are there
additional
contributions from this source?  Thinking back
to the graphical origin of structure function evolution [\wein ], 
we recognize
that such contributions 
originate from the soft side of the cut process, and thus
that in calculating the effect of virtual heavy particles
one would meet factors of $p^2/m^2$,
where $p^2$ is a typical hadronic momentum, making it relatively negligible.  

3.  A larger interference 
effect might be induced from 
exchange of an additional neutral gauge boson, say Z$^\prime$.  Such
gluon-Z$^\prime$ interference is somewhat analogous to Z-$\gamma$
interference
below threshold in e$^+$e$^-$ annihilation, with the following important
difference:
since
Z$^\prime$ unlike the gluon is a color singlet, 
interference effects only arise from crossed channels.  The analysis
of this case is worthwhile but 
much more complicated, and will not be undertaken here.

{\bf Acknowledgments}

P.K. would like to thank Peter Cho for helpful discussions.  
We also want to thank Chris Kolda for helpful comments on the manuscript.

{\bf Note added 5/21/96}: J. Ellis and D. Ross, in hep-ph/9604432,
have significantly extended the calculations reported here.  
We thank
them for pointing out some minor numerical errors in the original
version of this note.

\refout

\figout

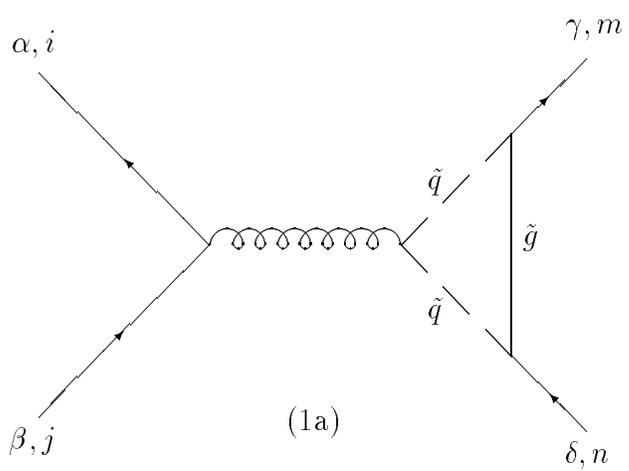
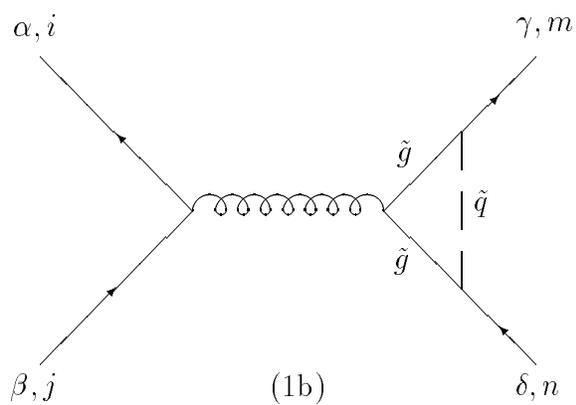
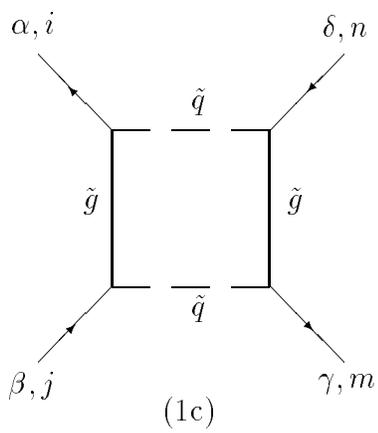
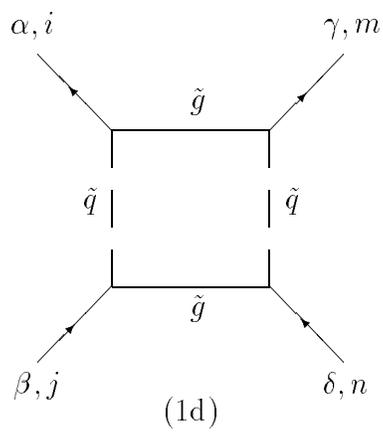
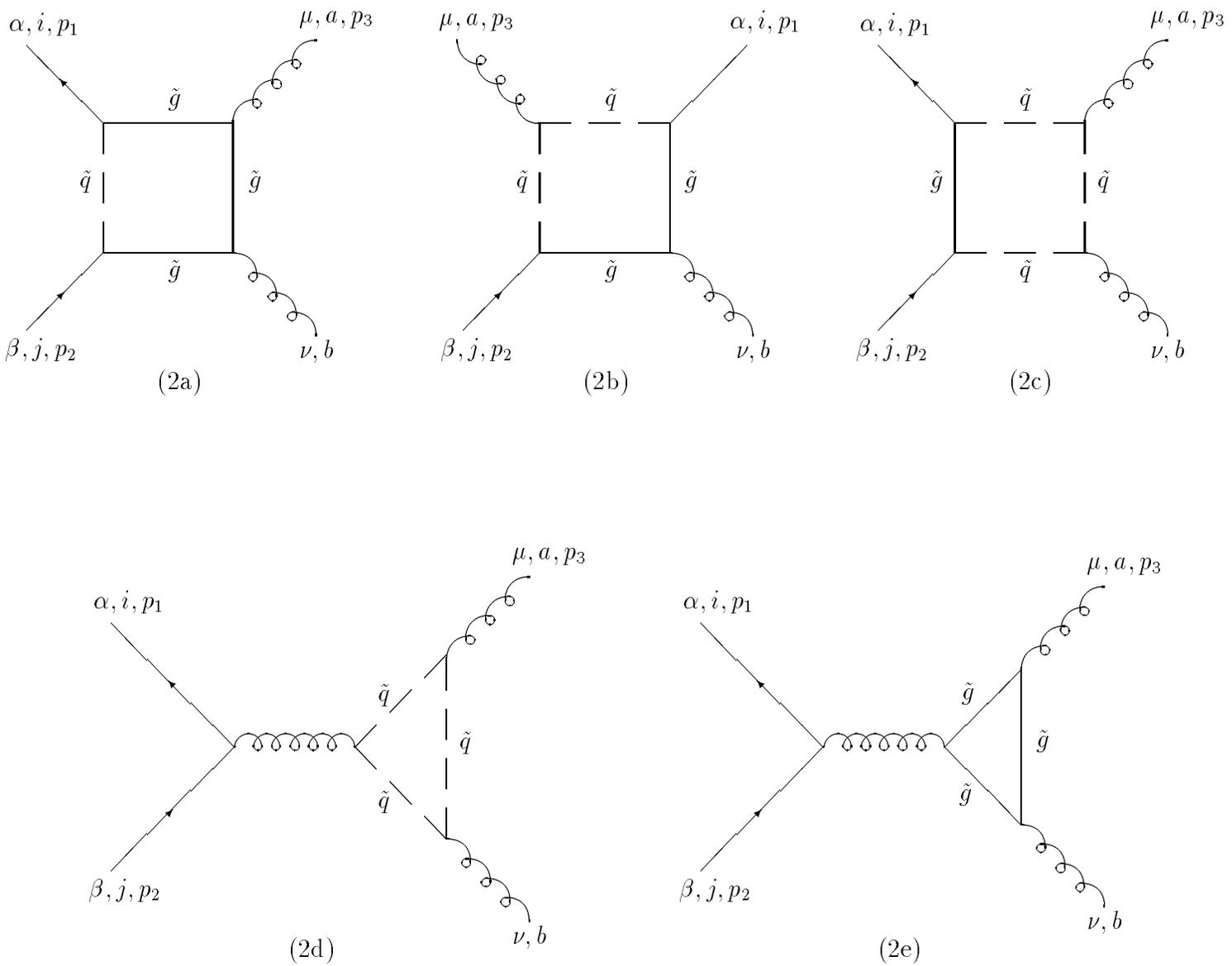
\end